\documentclass[12pt,aps,prd,preprint,preprintnumbers,tightenlines,twocolumn,superscriptaddress, showpacs,nofootinbib]{revtex4}

\usepackage{epsfig,latexsym,cancel,amssymb,amsmath}
\usepackage{hyperref}      
\usepackage{graphicx}
\usepackage{epstopdf}
\usepackage{color}
\usepackage{bm}
\usepackage{ulem}

\def\beq{\begin{equation}}
\def\eeq{\end{equation}}
\def\bea{\begin{eqnarray}}
\def\eea{\end{eqnarray}}

\def\refsec#1{section \ref{sec:#1}}

\begin{document}
\preprint{MSUHEP-130605}
\preprint{ NSF-KITP-13-105}
\title{Discriminating Higgs production mechanisms using jet energy profiles}

\author{Vikram Rentala}
\email{rentala@pa.msu.edu}
\affiliation{Department of Physics \& Astronomy, Michigan State University, E. Lansing, MI 48824, USA}

\author{Natascia Vignaroli}
\email{vignaroli@pa.msu.edu}
 \affiliation{Department of Physics \& Astronomy, Michigan State University, E. Lansing, MI 48824, USA}

\author{Hsiang-nan Li}
 \email{hnli@phys.sinica.edu.tw}
 \affiliation{Institute of Physics, Academia Sinica, Taipei, Taiwan 115, Republic of China}
 \affiliation{Department of Physics, National Cheng-Kung university, Tainan, Taiwan 701, Republic of China} \affiliation{Department of Physics, National Tsing-Hua university, Hsin-Chu, Taiwan 300, Republic of China}

\author{Zhao Li}
\email{zhaoli@ihep.ac.in}
  \affiliation{Institute of High Energy Physics, Chinese Academy of Sciences, Beijing 100049, China}

\author{C.-P. Yuan}
\email{yuan@pa.msu.edu}
\affiliation{Department of Physics \& Astronomy, Michigan State University, E. Lansing, MI 48824, USA}


\begin{abstract}
We present a new tool for precision measurements of the Higgs boson production mechanisms at the LHC. We study events with a Higgs boson produced with two forward jets. Even with fairly stringent cuts, one expects a significant contamination of gluon fusion (GF) in addition to vector-boson fusion (VBF) in the event sample. 
By measuring the jet energy profile of the most central jet, we find that SM production can be distinguished from either pure VBF or pure GF at the $5\sigma$ level with $100$ fb$^{-1}$ of luminosity at the 14 TeV LHC. Moreover, this discrimination technique can be used to validate or rule out new physics models that predict similar observable branching fractions as the $125$ GeV SM Higgs but have different production mechanisms.

\end{abstract}

\maketitle

\section{Introduction}
The discovery of a 125 GeV resonance by the ATLAS and CMS collaborations \cite{Aad:2012tfa,Chatrchyan:2012ufa} has raised many interesting questions about the nature of electroweak symmetry breaking and the Higgs mechanism. While we are still in the relatively early stages of collecting data about this particle, its properties appear to be consistent with that of a Standard Model (SM) Higgs boson. Given the apparent lack of evidence of weak-scale supersymmetry or compositeness, if this particle's properties continue to agree with those of a SM Higgs boson after the accumulation of even more data, then the problems of naturalness and hierarchy come to the forefront.

It is therefore crucial to measure every aspect of the purported Higgs boson in the coming years at the LHC and possibly at future colliders. Currently, most techniques have focussed on a global fit of the decay modes of the Higgs boson in order to determine its coupling to SM particles \cite{ATLAS-CONF-2013-034,CMS-PAS-HIG-13-005,H-global-fit}. In all studies of the Higgs decay modes, the experimental measurement is strictly speaking a product of the production cross-section times the decay branching ratio to a particular mode. This limits the discriminating power of using only the branching ratios to compare consistency with the SM Higgs. Moreover, decays to final states involving gluons (about 80$\%$ of the time) or light quarks cannot directly be tested due to the large QCD background.

At the LHC, the SM Higgs boson, with a mass around 125 GeV, has two dominant production mechanisms. The first is gluon fusion (GF) and the second is vector boson fusion (VBF). Thus, a verification of the Higgs production modes allows us to probe the Higgs coupling to gluons and weak bosons independently of the decay branching ratios.

Currently, attempts to separate the GF contribution from the VBF contribution to Higgs production rely on imposing kinematic cuts on the inclusive process $p p \rightarrow H + 2 \textrm{ jets}$ with $ H \rightarrow \gamma \gamma $. At least two hard jets separated by a large rapidity gap are required. However, in spite of these cuts, there is usually a sizeable contamination ($\sim 20\%)$ from gluon fusion and also an $\mathcal{O}(1)$ background contribution when looking for the Higgs decaying to two photons. In order to determine the gluon and the weak bosons couplings to the Higgs, one needs to disentangle the GF contribution from this sample.

We make the simple observation that the jets associated with VBF are initiated by quarks at the parton level, whereas the jets associated with GF are dominantly gluon-like. Thus, if we can measure the ratio of gluon to quark jets in the data (by considering the more central jet of the two leading $p_T$ jets, say),
 we can effectively measure the VBF production and GF cross-sections independently.

The main goal of this work is to demonstrate the effectiveness of separating the GF from the VBF contribution of SM Higgs production by examining the average energy profile of the more central jet. We construct a discrimination variable $f_V$, the fraction of VBF produced Higgs bosons in a given sample, and explain how to measure it. One could in principle look at any sample of $H + 2 \textrm{ jets}$ (not just in the above kinematic regime) and measure $f_V$ to check consistency with the SM prediction. This variable can be used \textit{not just to verify the Standard Model but to discriminate models of new physics} that predict different Higgs production mechanisms. We remark that although the discriminating variable $f_V$ is treated independently in this paper, one should include it in a global analysis to get even stronger constraints on Higgs couplings to gluons and vector bosons.

This paper is organized as follows: In \refsec{KS}, we describe the current discriminating power of Higgs production mechanisms at the LHC using conventional kinematic variables. In \refsec{jd}, we discuss the use of jet energy profiles to discriminate quark from gluon jets. In \refsec{procedure}, we describe the procedure to measure these jet energy profiles in Higgs production. In \refsec{fvdescr}, we describe how to construct the discriminating variable $f_V$. We simulate trials of LHC data to estimate the errors on the jet energy profiles and on $f_V$. The results of this numerical analysis are shown in \refsec{results}. The numerical results (with error bars) enable us to estimate the luminosity needed to discriminate SM Higgs production from pure VBF production or pure GF production at the $5\sigma$ level or more using 100 fb$^{-1}$ of luminosity at the 14 TeV LHC.
In \refsec{background}, we comment on the role of background on our analysis. While a full background study is beyond the scope of this work, we make some estimates of the effect of error from background on the discriminating power of our technique. Finally, we conclude in \refsec{conclusions} and make some suggestions for future applications of jet energy profiles to new physics searches.

\section{Kinematic Separation}
\label{sec:KS}
We consider the process $p p \rightarrow H + 2 \textrm{ jets}$ with $ H \rightarrow \gamma \gamma $. Based on the CMS analysis with di-jet tag in \cite{Chatrchyan:2012ufa}, we will consider the \textbf{tight} cut selection, in which the two photons satisfy the following requirements on their transverse momenta ($p_T$) and pseudorapidity ($\eta$):
\begin{equation}
p^{\gamma_1}_T> m_{\gamma\gamma}/2, \quad p^{\gamma_2}_T> m_{\gamma\gamma}/4, \quad |\eta_\gamma|<2.5,
\end{equation}
where $m_{\gamma\gamma}$ denotes the invariant mass of the two photons. The two jets satisfy:
\begin{align}
\begin{split}
& p^{j_1}_T> 30 \ \text{GeV}, \quad  p^{j_2}_T> 30 \ \text{GeV}, \quad |\eta_j|<4.7, \\
& \quad \Delta\eta_{jj}>3.5,
\end{split}
\end{align}
where $\Delta\eta_{jj}$ is the rapidity separation between the two jets. The following requirement on the invariant mass of two leading jets is also imposed:
\begin{equation}
M_{jj}> 500 \, \textrm{GeV}.
\end{equation}

For our analysis, we will also identify a second category of cuts, where we lower the cut on $M_{jj}$ to:
\beq
M_{jj}>250 \, \textrm{GeV} \ .
\eeq

The following conditions are applied to further reduce the background and enhance selection of VBF-like events: the difference between the average pseudorapidity of the two jets and
the average pseudorapidity of the diphoton system is required to be less than 2.5, and the difference
in azimuthal angle between the diphoton system and the dijet system is required to be greater
than 2.6 radians \cite{Chatrchyan:2012ufa}.

As mentioned in the introduction, in spite of these cuts, there is usually a sizeable contamination $\sim\mathcal{O}(20\%)$ from gluon fusion and also an $\sim\mathcal{O}(100\%)$ background contribution when looking for the Higgs decaying to two photons.

\begin{figure}
\includegraphics[width=0.4\textwidth]{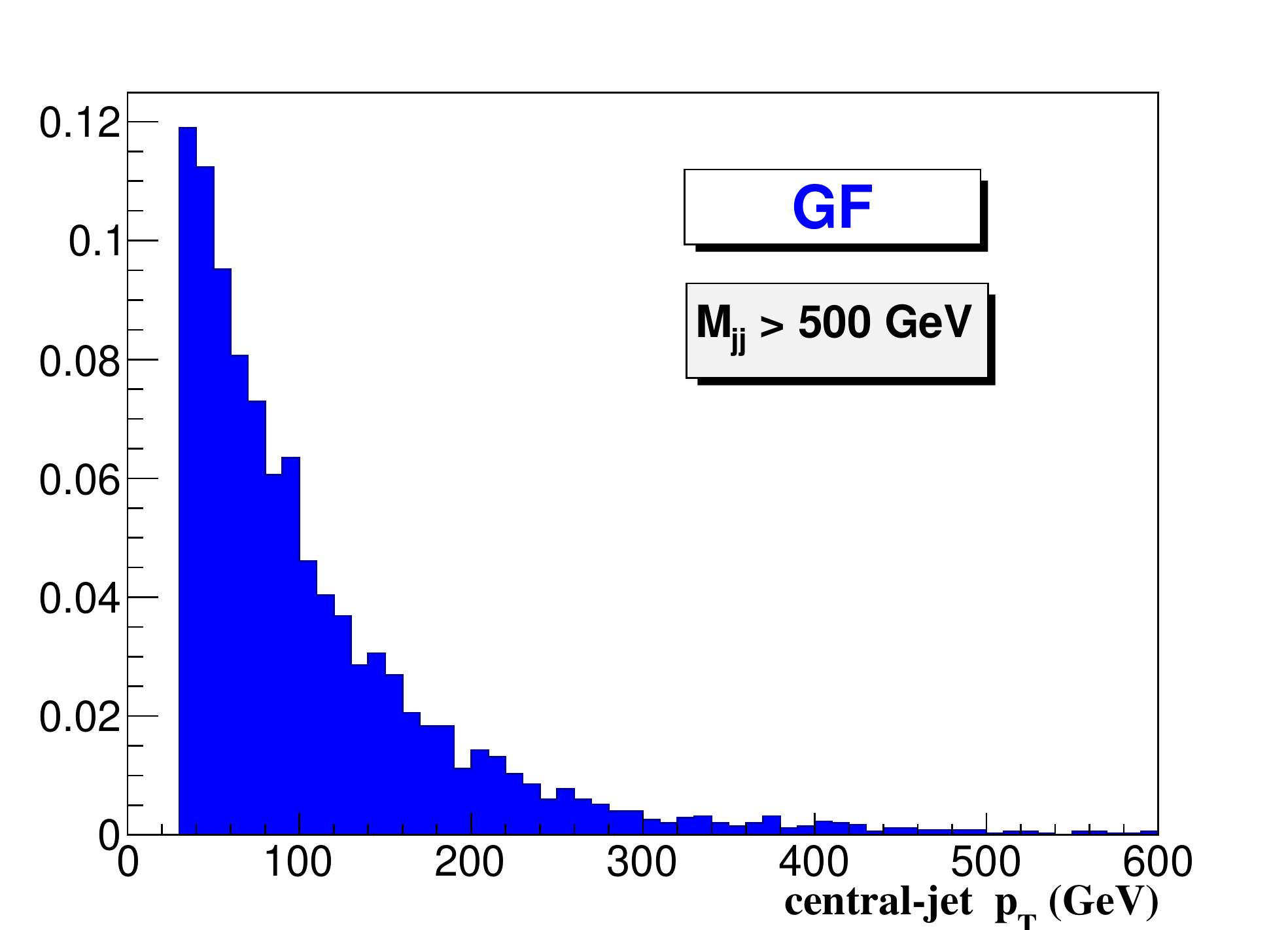}
\includegraphics[width=0.4\textwidth]{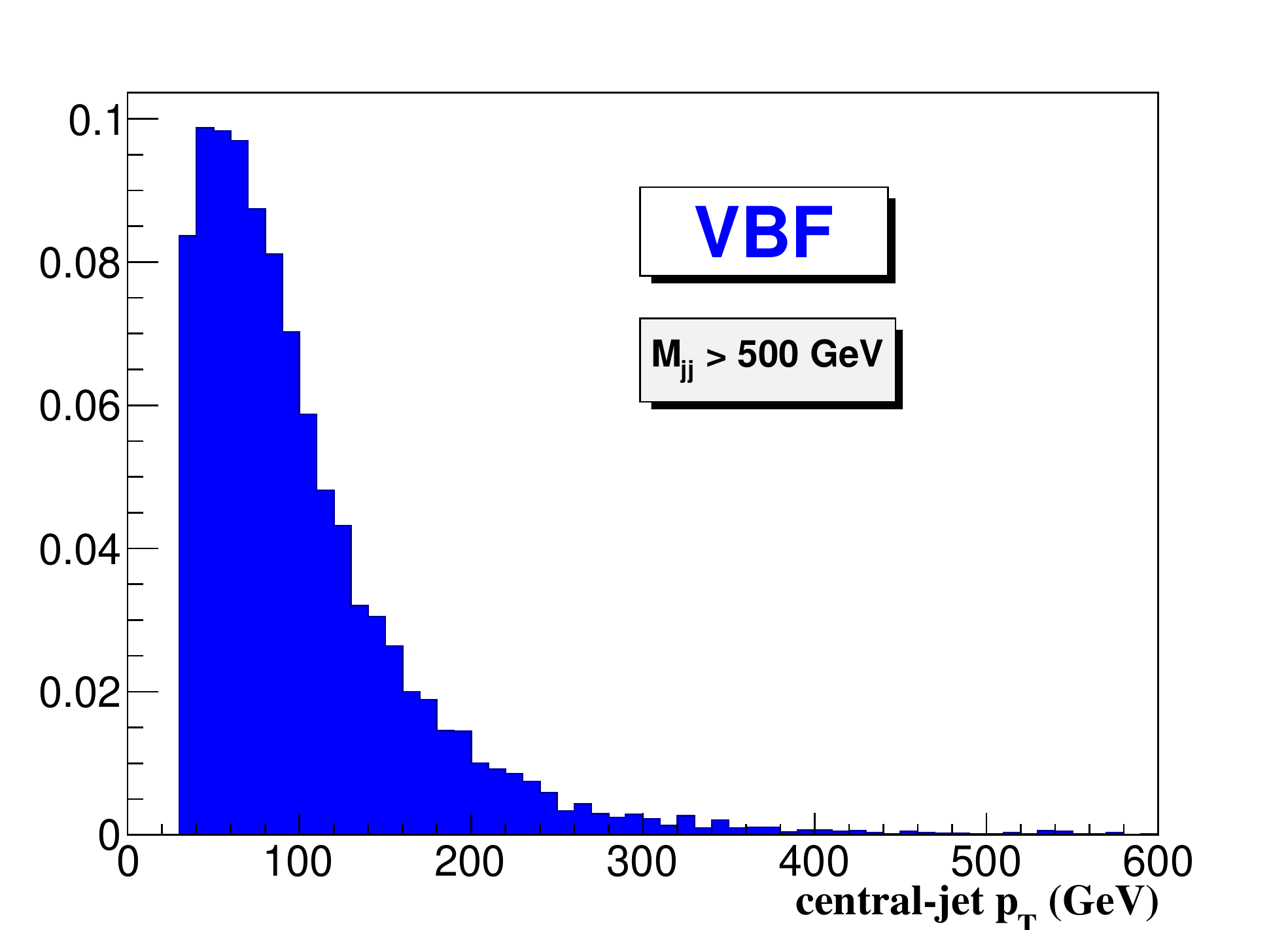}
\caption{{\small Normalized $p_T$ distribution of the central jet for GF (upper panel) and for VBF (lower panel) in $H + 2 \textrm{ jets}$ events passing the tight selection cuts with $M_{jj}>500$ GeV. 
}}
\label{fig:pt}
\end{figure}

As a first attempt to separate the GF and VBF contributions in this regime, one could try to use kinematic discriminants. 
Fig. \ref{fig:pt} shows the $p_T$ of the central jet for GF and for VBF. Both graphs have been normalized to have the same area.

We point out two interesting features of the $p_T$ spectra.
\begin{enumerate}
\item Towards the lower $p_T$ region, VBF jets are expected to typically have $p_T \sim M_W/2$, where $M_W$ is the $W$-boson mass. However, for GF, the associated jets have a lower $p_T$ spectrum.
\item In the high $p_T$ region the jets from GF have a more slowly decaying tail compared to the jets from VBF. This can be understood from the large invariant mass cut imposed on the two leading jets. The large pseudorapidity separation for jets in VBF ensures a large invariant mass (even for lower $p_T$ jets), however GF events are required to have harder jets in order to generate a large invariant mass.
\end{enumerate}

\begin{figure}
\includegraphics[width=0.4\textwidth]{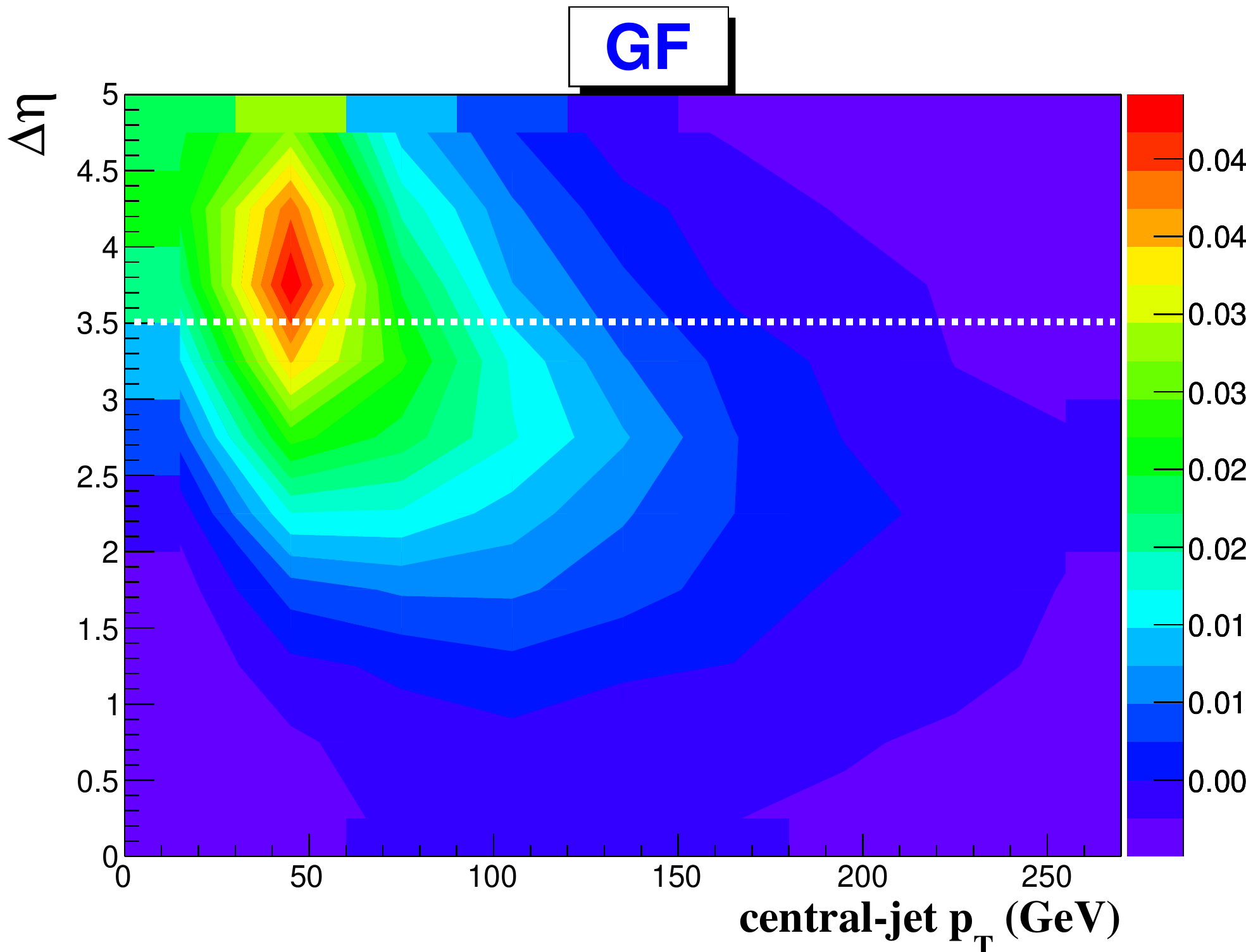}
\includegraphics[width=0.4\textwidth]{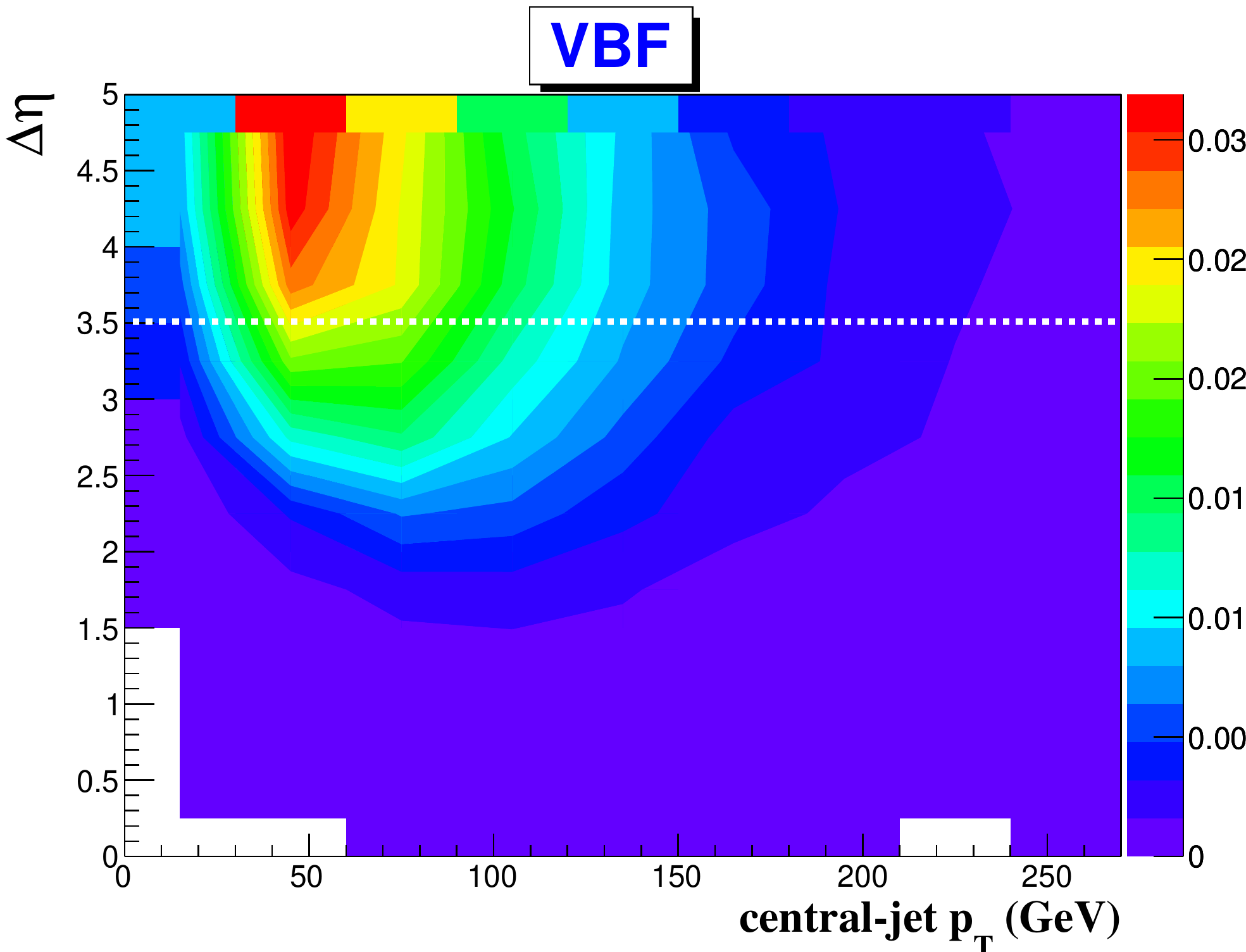}
\caption{{\small $p_T$ of the central jet vs $\Delta\eta$ of the two jets for GF (upper panel) and for VBF (lower panel) in $H + 2 \textrm{ jets}$ events, when only mild cuts on jets are applied. The dotted white line shows the value of the cut on $\Delta\eta$ applied in the analysis.}}
\label{fig:ptVseta}
\end{figure}

Fig. \ref{fig:ptVseta} shows a contour plot of the $p_T$ of the central jet versus the rapidity separation $\Delta \eta$ of the two jets. We can see the larger pseudorapidity separation for VBF, as alluded to in point 2 above.

\begin{figure}
\includegraphics[width=0.4\textwidth]{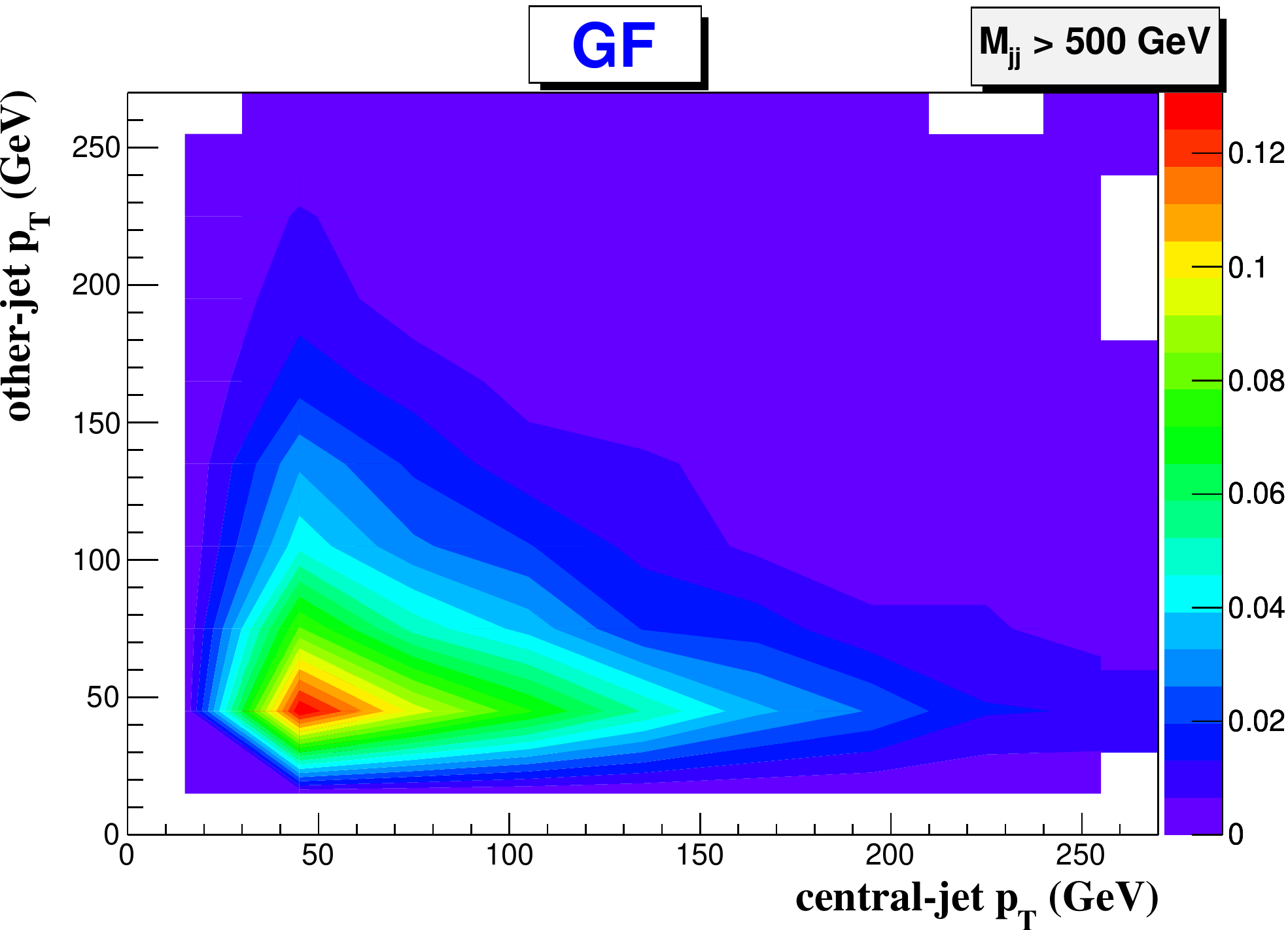}
\includegraphics[width=0.4\textwidth]{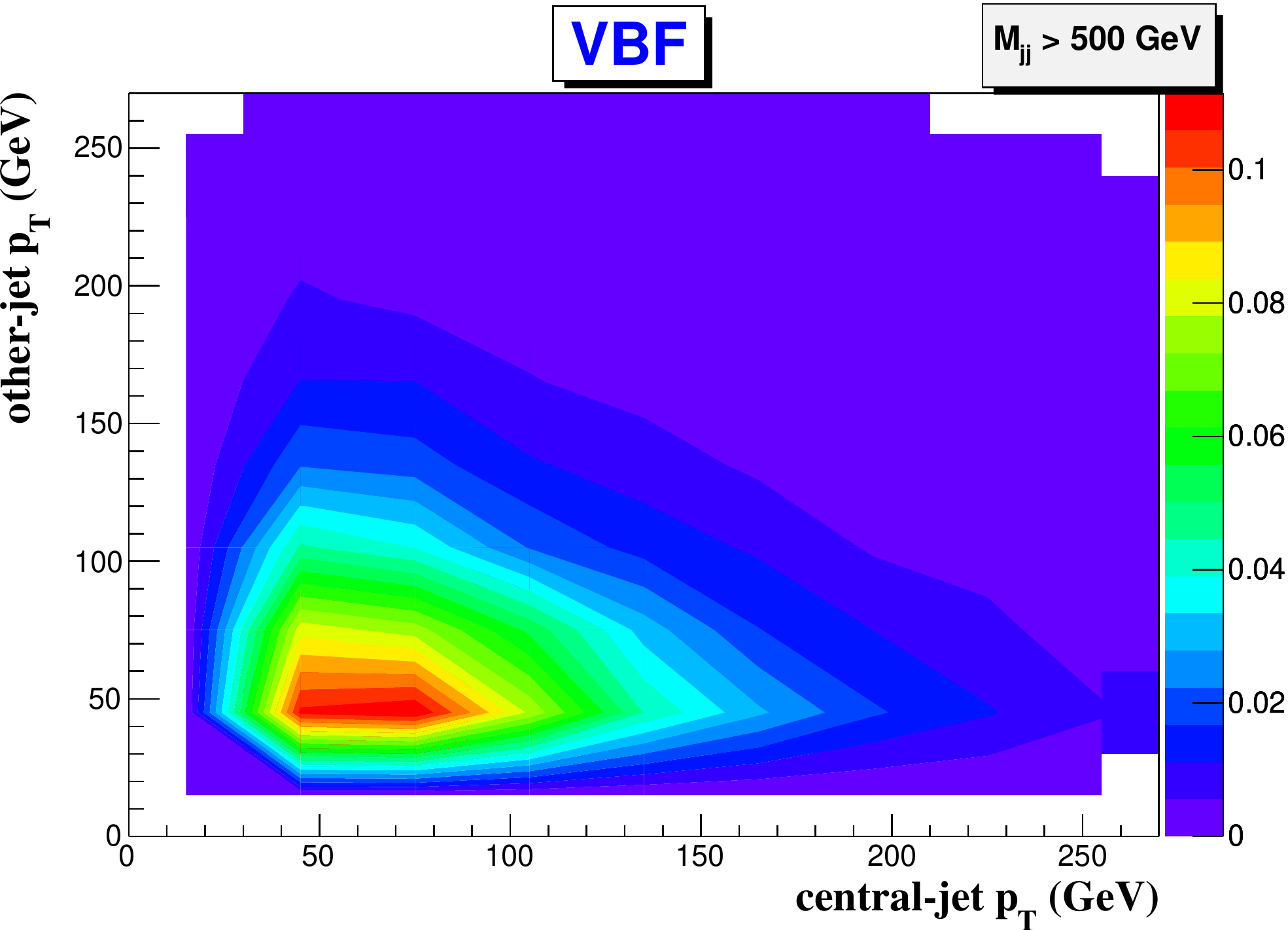}
\caption{{\small $p_T$ of the central jet vs $p_T$ of the other jet for GF (upper panel) and for VBF (lower panel) in $H + 2 \textrm{ jets}$ events passing the tight selection cuts with $M_{jj}>500$ GeV.}}
\label{fig:ptVspt}
\end{figure}

Fig. \ref{fig:ptVspt} is a contour plot showing the $p_T$ of the central jet versus the $p_T$ of the less central jet for GF and VBF. We see that more often than not the leading $p_T$ jet is the same as the central jet.

We summarize this section by noting that although the kinematic distributions highlight interesting differences between GF and VBF production mechanisms, the distributions look quite similar once tight kinematic cuts are applied. This shows the difficulty of trying to further separate GF and VBF by simply strengthening the kinematic cuts. New discriminating tools, such as the study of jet shapes described in this paper can provide a more effective way to achieve this separation.

\section{A New Discriminant: Jet Energy Profiles}
\label{sec:jd}
In this paper we will use jet energy profiles (JEPs) to statistically distinguish quark jets from gluon jets. For a jet of size $R$, the (integrated) JEP, $\psi(r)$, is defined as the fraction of jet transverse energy in a cone of size $r(<R)$ concentric to the original jet cone,
\begin{equation}
\label{eq:psi}
\psi(r) = \frac{ \sum\limits_{r'<r}^{} p_T(r')}{\sum\limits_{r'<R}^{} p_T(r')}.
\end{equation}
These JEPs were measured at the Tevatron CDF experiment \cite{Acosta:2005ix}.
More recently, JEPs have been measured at ATLAS \cite{Aad:2011kq} and at CMS \cite{CMS-PAS-QCD-10-013}.
It was found that the experimental data can be described by a carefully tuned event generator, PYTHIA Tune A \cite{Sjostrand:2007gs,pythiaA}.

Gluon initiated jets are expected to spread more due to more radiation and thus have a slowly rising JEP. Quark initiated jets on the other hand radiate less, and thus accumulate a larger fraction of their energy for fairly small $r$ and have a quickly rising JEP. Both profiles approach unity as $r\rightarrow R$ because of the normalization.

One can calculate these profiles in perturbative QCD (pQCD) for quark or gluon initiated jets. It has been found that the next-to-leading order (NLO) profiles overshoot the data \cite{Li:2011hy} at small $r$, where the logarithmic corrections $\alpha_s\log(R/r)$ are important and need to be resummed. The invalidity of fixed-order analysis of jet observables motivates the recent development of the resummation formalism: an improved next-to-leading-logarithm (NLL) resummation calculation of the JEP was performed in \cite{Li:2011hy,Li:2012bw}, which organizes terms of the form $\alpha^n_s \left (\log(R/r) \right )^{2n}$ and $\alpha^n_s \left (\log(R/r) \right )^{2n-1}$ to all orders in $\alpha_s$. The JEP also changes with the total $p_T$ of the jet due to QCD scaling violation. Because the running coupling constant decreases with $p_T$, the resummation effect is expected to be minor at high $p_T$. This is the reason why the resummation predictions approach the NLO ones as $p_T$ increases \cite{Li:2011hy,Li:2012bw}. Very good agreement was found between the data and the resummation calculation for a wide range of $p_T$. It was also theoretically confirmed that a gluon jet is broader than a quark jet with the same $p_T$.

However at (large) $p_T >200$ GeV, some deviations were observed at low $r \sim 0.1$, where the resummation predictions fall a bit below the data. This deviation may be attributed to non-perturbative effects from hadronization and underlying events, or to higher-power effects in the resummation formalism \cite{Li:2012bw}. For example, the phase space of the soft gluons that contribute to the anomalous dimension of the resummation was overestimated, and the overestimate, being proportional to $r$ \cite{Li:2012bw}, was regarded as a power correction to the energy profile. The energy profile is normalized to unity at $r=R$, so the overestimate actually causes suppression of the distribution at small $r$, explaining the slight drop of the resummation predictions in comparison with the data. The overestimate is more pronounced at larger $p_T$ due to the narrowness of the jet, explaining why the deviation becomes more obvious at high $p_T$. These uncertainties should be kept in mind when comparing our predictions to the data.

As discussed in \cite{Li:2011hy,Li:2012bw}, the factorization of the soft gluons that contribute to the anomalous dimension of the resummation for an energetic jet can be regarded equivalently as associating these soft gluons with the clustered jet. This prescription is exactly what was adopted in the anti-$k_t$ algorithm. In order to match theoretical calculations to the data consistently, we will consider anti-$k_t$ jets \cite{Cacciari:2008gp} with a cone size of $R = 0.7$.

An important point, about comparing the theoretical prediction to the experimental results, is that the theory calculation involves some (arbitrary) scale parameters which are introduced into pQCD calculations to estimate the effect of the yet-to-be calculated sub-leading logarithmic contributions.
Hence, the variation in the theory prediction for different values of the scale parameters can be taken as the theoretical error in our calculation. It is expected that the experimental errors at the LHC will be  smaller than this theoretical uncertainty.

Thus, before applying our theory prediction to the energy profile of the jets produced in association with the Higgs boson observed at the LHC, we could test our prediction with precision experimental data from a known process, such as  $Z + 2 \textrm{ jets}$ events, in which the kinematics of the observed jets could be chosen to be similar to those in the $H + 2 \textrm{ jets}$ events. In other words, we could use the comparison to the $Z +  2 \textrm{ jets}$ data to calibrate our prediction for the JEP of the jets associated with the Higgs boson produced at the LHC.

\section{Analysis Procedure}
\label{sec:procedure}
We now explain how to obtain a numerical prediction for the JEPs for the central jet in $H + 2 \textrm{ jets}$ events with $H \rightarrow \gamma\gamma$ in the Standard Model. We also consider two hypothesis models for comparison. 
The first is pure VBF production and the second is pure GF production. For these test scenarios, we will rescale the total cross-sections to agree with the SM Higgs boson production rates after imposing the relevant kinematic cuts. 

First, we simulate $H + 2 \textrm{ jets}$ events using MadGraph v5 \cite{Alwall:2011uj}\footnote{Specifically, we use the so-called ``Higgs Effective Field Theory'' model with a 125 GeV Higgs and with the corresponding LO width.}. The results of our leading-order (LO) simulations for the 8 TeV LHC are compared to the CMS results in \cite{Chatrchyan:2012ufa} for the tight selection cuts with $M_{jj} > 500 $ GeV, introduced in Sec.~\ref{sec:KS}. In order to incorporate higher-order corrections to the cross-section, we use a ``$k$-factor'' correction, that rescales the simulated MadGraph cross-section to agree with the CMS results\footnote{As a sanity check, we also find good agreement with the CMS data using the same $k$-factors but looking in the ``loose'' cut region identified in \cite{Chatrchyan:2012ufa}.}.

We will use these same $k$-factors, quoted in Table \ref{tab:8tev-xsec}, for the simulation at 14 TeV. Table \ref{tab:14tev-xsec} shows the expected cross-sections at the 14 TeV LHC for tight selection  cuts with  $M_{jj} > 500 $ GeV, and for the case $M_{jj}> 250$ GeV. Our analysis will be efficient for a sufficiently high number of events, of the order of one-hundred, and we will thus be focused on the 14 TeV center-of-mass energy, expected at the next stage of operation of the LHC.

\begin{table}
\begin{tabular}{|c|cc|}
\cline{2-3}
\multicolumn{1}{c}{} & \multicolumn{2}{|c|}{ $\mathbf{M_{jj} > 500}$ \bf{GeV} } \\[0.02cm]
\hline
\textsf{8 TeV} & GF & VBF  \\[0.05cm]
\cline{2-3}
CMS & 23\% & 77\%\\[0.05cm]
& 0.11 fb & 0.38 fb  \\[0.05cm]
\hline
 $K^{\textrm{CMS}}_f$ & 1.6 & 1.2\\
 \hline
\end{tabular}
\\[0.5cm]
\caption{{\small CMS cross-sections  at the 8 TeV LHC using tight cuts and the corresponding compositions of VBF and GF to the total SM rate \cite{Chatrchyan:2012ufa}. The factor, $K^{\textrm{CMS}}_f$, is the correction factor needed to rescale the MadGraph cross-sections to agree with the CMS data.}}\label{tab:8tev-xsec}
\end{table}

\begin{table}
\begin{tabular}{|c|cc|cc|}
\cline{2-5}
\multicolumn{1}{c}{} & \multicolumn{2}{|c}{$\mathbf{M_{jj} > 500}$ \bf{GeV}}& \multicolumn{2}{|c|}{$\mathbf{M_{jj}>250}$ \bf{GeV}}\\[0.02cm]
\hline
\textsf{14 TeV} & GF & VBF & GF & VBF \\[0.05cm]
\hline
MG $\times$ $K^{\textrm{CMS}}_f$ & 32\% & 68\% & 38\% & 62\% \\[0.05cm]
& 0.57 fb & 1.2 fb & 0.88 fb & 1.4 fb \\
\hline
\end{tabular}
\caption{{\small SM expected cross-sections at the 14 TeV LHC, using tight cuts with $M_{jj}>500$ GeV and with $M_{jj}>250$ GeV.}}\label{tab:14tev-xsec}
\end{table}

\subsection{Calculating the JEPs}
Now, we are ready to calculate the theoretical predictions from pQCD for the JEP of the central jet using the formulae derived in \cite{Li:2011hy, Li:2012bw}. We take jet four-momenta from MadGraph and convolve with the resummed jet functions to arrive at the numerical JEPs.

\begin{figure}
\includegraphics[width=0.45\textwidth]{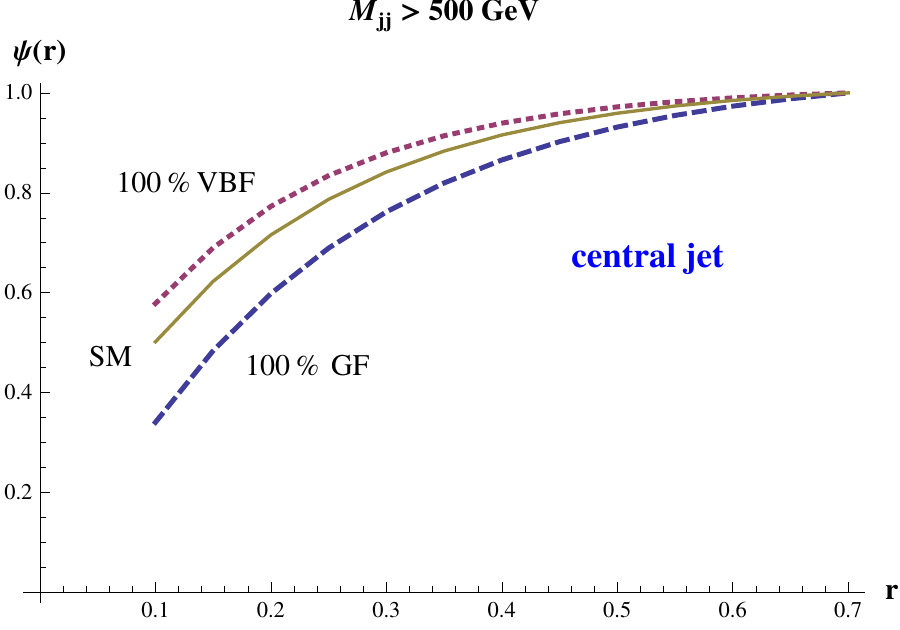}
\includegraphics[width=0.45\textwidth]{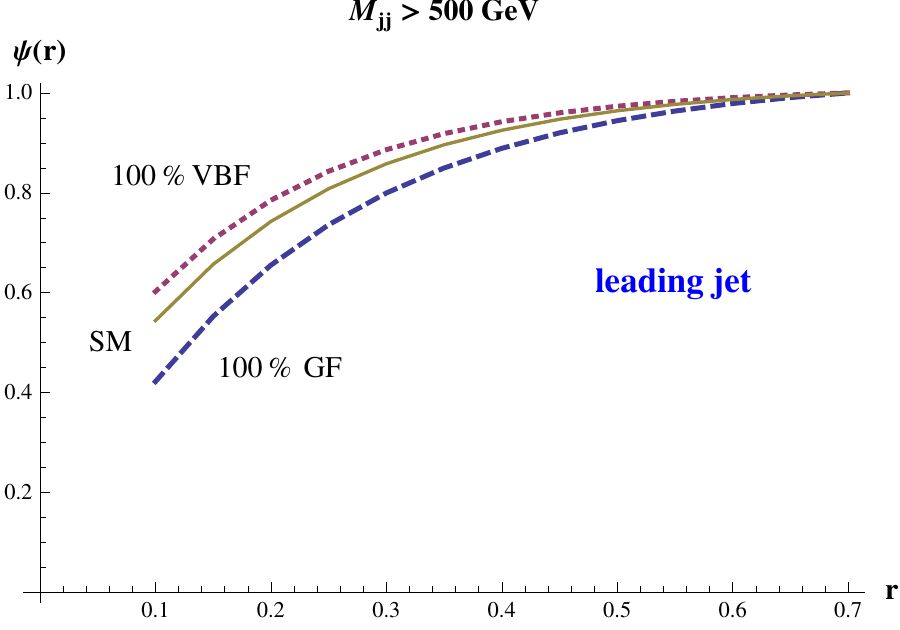}
\caption{{\small Energy profile of the most central jet (upper panel) and of the leading $p_T$ jet (lower panel) in the $H+ 2\textrm{ jets}$ events that have passed the tight cuts with $M_{jj}>500$ GeV, in the SM and in the hypothetical cases of a Higgs produced via pure GF and via pure VBF. }}
\label{fig:tight-central-leading}
\end{figure}

The JEP of the most central jet in the $H + 2 \textrm{ jets}$ events passing the cuts with $M_{jj} > 500$ GeV is shown in Fig. \ref{fig:tight-central-leading} (upper panel) for pure VBF production, pure GF production and for the SM. The lower panel of Fig. \ref{fig:tight-central-leading} shows, for comparison, these same profiles when we consider the leading $p_T$ jet, instead of the most central jet. We see that the discriminating power is somewhat reduced in this case. This retroactively justifies our use of the central jet for discrimination.

\begin{figure}
\includegraphics[width=0.45\textwidth]{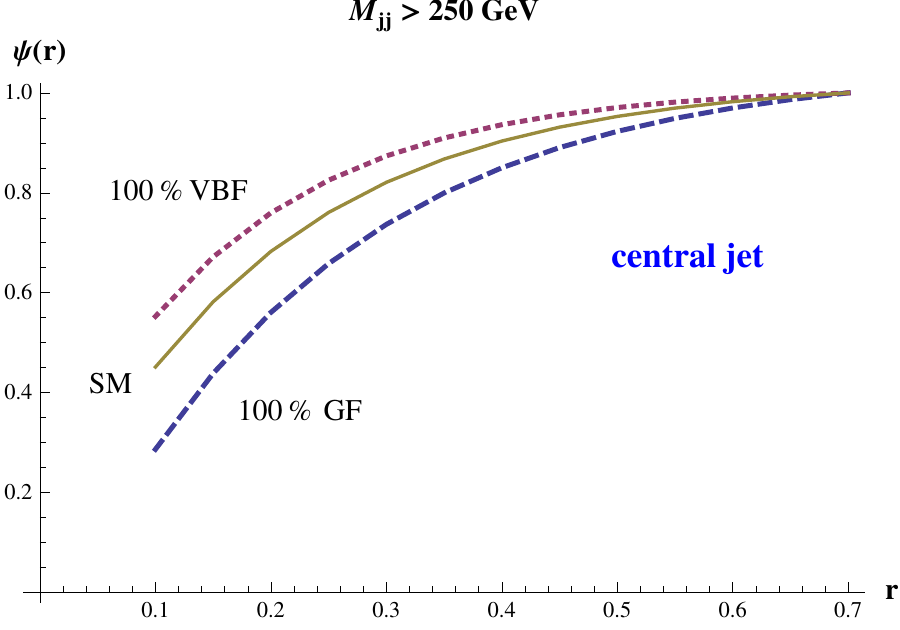}
\caption{{\small Energy profile of the most central jet in the $H+ 2\textrm{ jets}$ events which satisfy tight cuts with $M_{jj}>$ 250 GeV, in the SM and in the hypothetical cases of a Higgs produced via pure GF and via pure VBF. }}
\label{fig:loose}
\end{figure}

\begin{figure}
\includegraphics[width=0.45\textwidth]{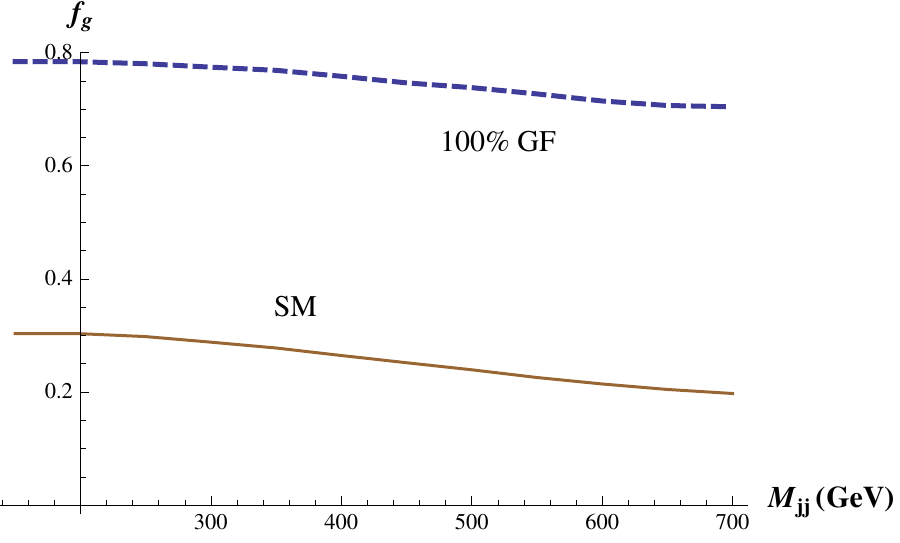}
\includegraphics[width=0.45\textwidth]{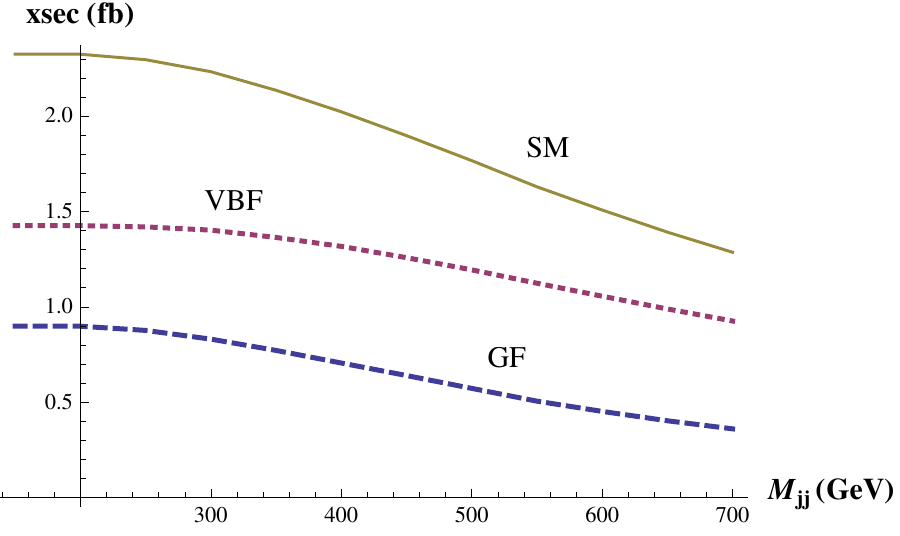}
\caption{{\small Upper Panel: Fraction of $H+ 2\textrm{ jets}$ events where the most central jet is a gluon, $f_g$, as function of the $M_{jj}$ cut (all the other cuts are the same as in the tight selection), in the SM and in the hypothetical case of a Higgs produced via pure GF. Lower Panel: Cross-sections as function of the $M_{jj}$ cut. }}
\label{fig:gluon-Mjj}
\end{figure}

We also observe a larger separation between the VBF, GF and SM profiles when the cut on $M_{jj}$ is lowered to $250$ GeV (Fig. \ref{fig:loose}). The reason for this effect is an increase of the fraction ($f_g$) of events where the most central jet is a gluon in pure GF (and in the SM) when the $M_{jj}$ cut is lowered (Fig. \ref{fig:gluon-Mjj}). However, for a milder $M_{jj}$ cut, we also expect a larger contamination from background; this effect will be discussed in Section \ref{sec:background}.

\subsection{Estimating the errors on the JEPs}
In order to determine how efficiently these profiles can be distinguished from each other, we need to estimate the statistical uncertainties on the theoretical profiles for a given integrated luminosity at the 14 TeV LHC. These statistical errors are obtained by studying the substructure of the reconstructed more central jet (of the two leading $p_T$ jets) in the full event sample including effects from parton showering and hadronization through Pythia v6.4 \cite{Sjostrand:2007gs} with the default tune.

We simulate Higgs +1,2,3 jets events with MadGraph, then we pass them to Pythia for showering and hadronization and we apply the MLM prescription \cite{Mangano:2006rw} for matching\footnote{We have used a cutoff scale QCUT=15 GeV and a xqcut=10 GeV scale. We refer the reader to \cite{Alwall:2007fs} for details on how to use matching inside MadGraph.}. Jets are reconstructed using SpartyJet \cite{Delsart:2012jm}, a wrapper for FastJet \cite{Cacciari:2011ma}, using the anti-$k_t$ algorithm with $R=0.7$. We first apply the selection cuts described in Sec. \ref{sec:KS} to the two leading jets in the final state\footnote{We find that the final cross-sections obtained in this manner are in good agreement (within 5\%) with the parton-level final cross-sections listed in Table \ref{tab:14tev-xsec}.}.

\begin{figure}
\includegraphics[width=0.45\textwidth]{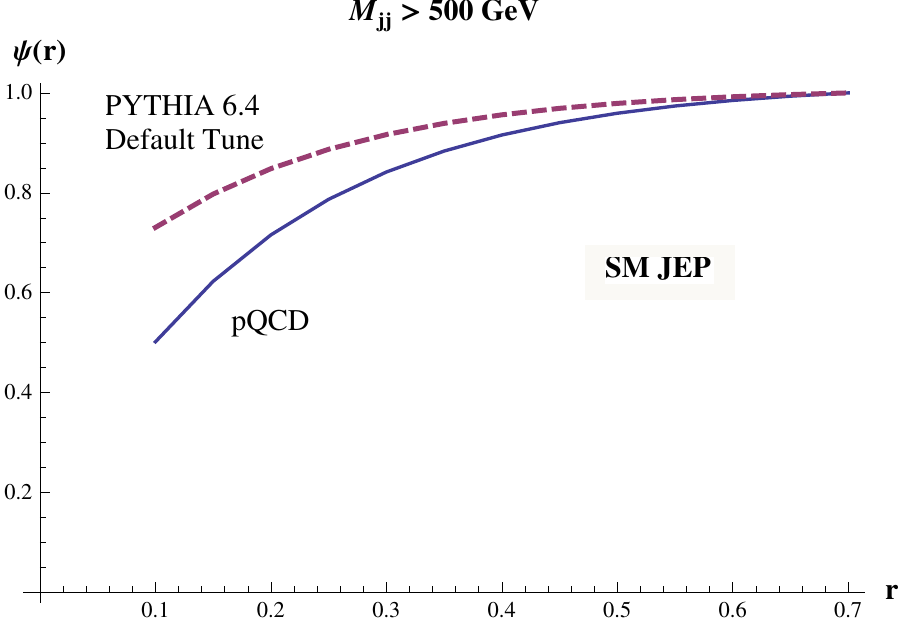}
\caption{Energy profile of the central jet for SM obtained by analyzing the jet substructure after Pythia v6.4 (default tune) showering, compared to the theoretical pQCD prediction using jet functions \cite{Li:2011hy,Li:2012bw}. }
\label{fig:jep-pythia}
\end{figure}

Next, using our Pythia event sample, we examine the central jets and, for a given sub-cone of size $r$, we calculate the mean of the integrated energy distribution, $\psi(r)$. In Fig. \ref{fig:jep-pythia} we show the energy profile of the central jet for SM events, obtained from Pythia showering, compared to that from the theoretical calculation. Notice the large difference between the theoretical and the Pythia prediction, which depends on the specific Pythia tune considered.

\textit{Due to this tune-dependence, we will rely on our theoretical calculations to determine the central value of the JEP, however we will use the Pythia results to estimate the errors on the JEPs.}

Using the same Pythia event sample, we can calculate the statistical variation on the sample-mean of the JEP, $\psi(r)$, for a given number of events. We find, as expected, that the variations follow Gaussian distributions and that the errors scale as the square root of the number of events.

\begin{figure}
\includegraphics[width=0.45\textwidth]{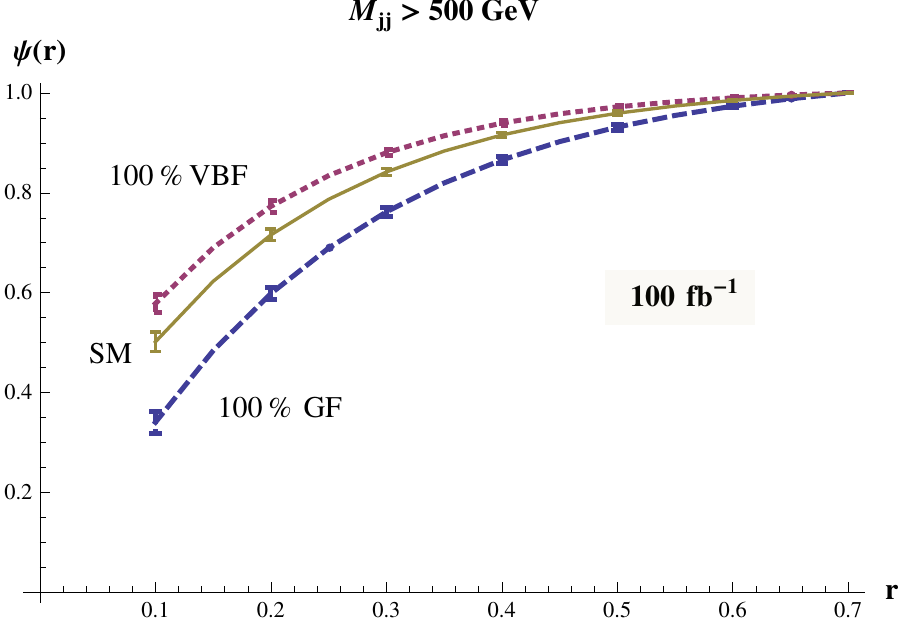}
\includegraphics[width=0.45\textwidth]{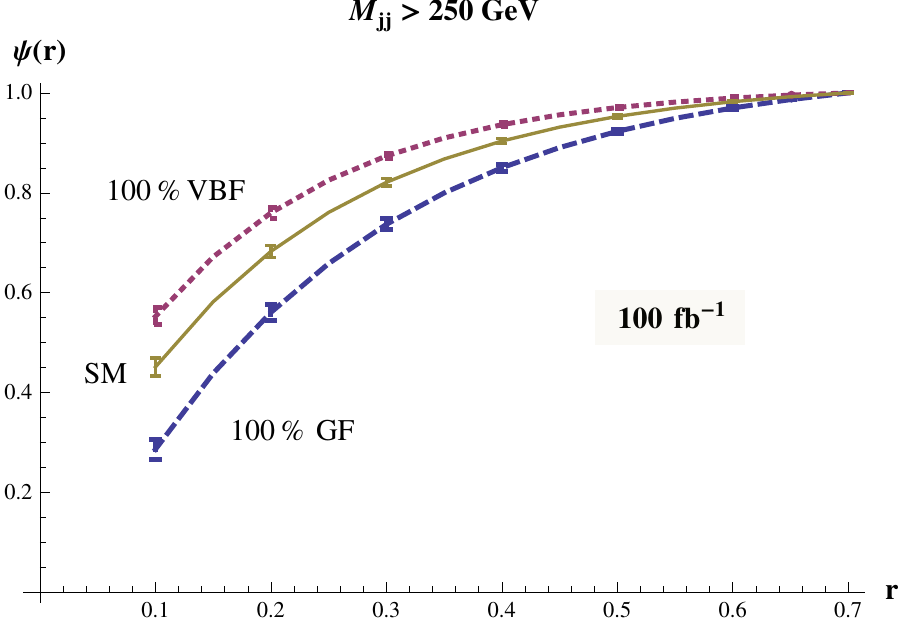}
\caption{{\small Energy profile of the most central jet in the $H+ 2 \textrm{ jets}$ events which satisfy tight cuts with $M_{jj}>500$ GeV (upper panel) and with $M_{jj}>$ 250 GeV (lower panel), in the SM and in the hypothetical cases of a Higgs produced via pure GF and via pure VBF. The statistical errors are derived from MadGraph + Pythia simulations by considering the number of events predicted for the SM with 100 fb$^{-1}$ of integrated luminosity at the 14 TeV LHC.} }
\label{fig:jep-final}
\end{figure}

Fig. \ref{fig:jep-final} shows our final result for the JEPs in the SM and in the hypothetical cases of a Higgs produced via pure GF and via pure VBF. The mean values of $\psi(r)$ are derived from the theoretical perturbative calculation, the error bars are the $1\sigma$ variations estimated from Pythia simulations by using the method explained above and considering the number of events predicted for the SM with 100 fb$^{-1}$ of data at the 14 TeV LHC.

\section{The discriminating variable $f_V$}
\label{sec:fvdescr}
JEPs can be well fitted by a two-parameter function of $(a, b)$,
\begin{equation}\label{eq:fit}
\psi(r) = \frac{1-b e^{-a r}}{1-b e^{- a R}}, \
\end{equation}
where $R = 0.7$ is the jet cone size. The values of $a$ and $b$ for SM, VBF and GF events are shown in Table \ref{tab:fitparams}. We will use these analytical expressions to approximate the numerical JEPs shown in Fig. \ref{fig:jep-final} in what follows.

\begin{table}
\begin{tabular}{|c|c|c|}
\cline{2-3}
\multicolumn{1}{c}{}  & \multicolumn{1}{|c}{$a$} &  \multicolumn{1}{|c|}{$b$} \\
\hline
SM & $ 5.3  $  &  $ 0.87$ \\
VBF & $ 6.0 $  & $ 0.83 $ \\
GF & $ 4.4 $  & $ 1.1 $ \\
\hline
\end{tabular}
\caption{ Fit parameters $a$ and $b$ for the central JEPs of SM, pure VBF and pure GF events using tight cuts with $M_{jj} > 250 $ GeV. The parameters are to be used in the analytic approximation, Eq. (\ref{eq:fit}), for the JEPs.}
\label{tab:fitparams}
\end{table}

We define a one parameter series of JEPs that linearly interpolate between the VBF and GF profiles,
\begin{align}
\label{eq:fV}
\begin{split}
\psi_{f_V} (r) & = f_V\, \psi_{\textrm{VBF}}(r) \\
& + (1-f_V)\, \psi_{\textrm{ GF} } (r) \ .
\end{split}
\end{align}
Here, $f_V$ is the parameter and $\psi_{\textrm{VBF}}(r)$ and $\psi_{\textrm{GF}}(r)$ are the pure VBF and pure GF profiles that are determined from the MadGraph simulation convolved with jet functions, as described in Sec. \ref{sec:procedure}.

For any experimentally measured JEP, we can perform a fit to the one-parameter family of curves and find the corresponding value of $f_V$. We see that $f_V$ has a clear physical meaning as the fractional amount of VBF contribution to $H + 2\textrm{ jet}$ production. Explicitly, $f_V = 1$ for pure VBF and $f_V = 0$ for pure GF production. The central values of $f_V$ for the SM for different cut choices are shown in Table \ref{tab:fVerrors}. These values agree with the expected VBF fractions from Table \ref{tab:14tev-xsec}, as they should.

Strictly speaking, the profiles are also functions of the jet $p_T$ as explained in Sec. \ref{sec:jd}. Hence, Eq. (\ref{eq:fV}) should be viewed as being valid only for a narrow $p_T$ band, and the parameter $f_V$ should also be regarded as a $p_T$-dependent parameter. In practice, we can ignore this subtlety for two reasons. Firstly, a factor-of-a-few change in $p_T$ is required to see appreciable changes in the profile. Secondly, by examining the $p_T$ distributions for the events under consideration (Fig. \ref{fig:pt}), we see that the bulk of the events are picked up in a fairly narrow $p_T$ band and thus we can safely ignore the $p_T$ dependence of Eq. (\ref{eq:fV}) in what follows.

\section{Results}
\label{sec:results}
Now using our simulated JEPs with error bars shown for 100 fb$^{-1}$ of integrated luminosity at the 14 TeV LHC (Fig. \ref{fig:jep-final}), we can translate the errors on the JEPs into errors on the measurement of the fitted $f_V$. The results are shown in Table \ref{tab:fVerrors} for a SM sample and pure VBF, pure GF samples. Note that the errors are Gaussian, and scale as the square root of the number of events (or equivalently the integrated luminosity).

\begin{table}
\begin{tabular}{|c|c|c|}
\hline
 \multicolumn{1}{|c|}{$f_V$} &\multicolumn{1}{c|}{$\mathbf{M_{jj} > 500}$ \bf{GeV} }& \multicolumn{1}{c|}{$\mathbf{M_{jj}>250}$ \bf{GeV}}\\
  \hline
SM & $0.68 \pm 0.05$  & $0.62 \pm 0.04$ \\
VBF & $1.00 \pm 0.04$  & $1.00 \pm 0.03$ \\
GF & $0.00 \pm 0.06$  & $0.00 \pm 0.05$ \\
\hline
\end{tabular}
\caption{ Fraction of VBF-like events ($f_V$) for SM, pure VBF and pure GF events with error bars shown for both tight cuts with $M_{jj}>500$ GeV and with $M_{jj}>250$ GeV. $f_V$ is determined by performing a fit to the one-parameter family of candidate JEPs defined in Eq. (\ref{eq:fV}).}
\label{tab:fVerrors}
\end{table}

From the results of Table \ref{tab:fVerrors}, we can determine our ability to discriminate SM events from a pure VBF sample or a pure GF sample. For 100 fb$^{-1}$ of integrated luminosity, we calculate the difference between the best-fit $f_V$ for VBF or GF and the best-fit $f_V$ for SM events and express the result in terms of standard deviation difference. Namely, we define the $\sigma$-level separation as:
\begin{equation}
\sigma^{VBF/GF}\equiv \frac{\left|f^{VBF/GF}_V-f^{SM}_V\right|}{\sqrt{\left(\sigma^{VBF/GF}_{f_V}\right)^2+\left(\sigma^{SM}_{f_V}\right)^2}}.
\end{equation}
For instance, a $2\sigma$ separation between VBF and the SM profiles indicates that the pure VBF hypothesis can be ruled out at the $2\sigma$ level. The results of this calculation for different dijet invariant mass cuts are shown in Table \ref{tab:fVsigma}.

\begin{table}
\begin{tabular}{|c|cc|cc|}
\cline{2-5}
\multicolumn{1}{c}{} & \multicolumn{2}{|c}{$\mathbf{M_{jj} > 500}$ \bf{GeV}}& \multicolumn{2}{|c|}{$\mathbf{M_{jj}>250}$ \bf{GeV}}\\ \cline{2-5}
 \multicolumn{1}{c}{} & \multicolumn{1}{|c}{GF} & \multicolumn{1}{c}{VBF} & \multicolumn{1}{|c}{GF}& \multicolumn{1}{c|}{VBF} \\[0.05cm]
\hline
$\sigma-$level & 8.7  & 5.0 & 9.7  & 7.6 \\
\hline
\end{tabular}
\caption{{\small Expected $\sigma-$level distinction between SM and pure GF or VBF event samples using 100 fb$^{-1}$ of luminosity at the 14 TeV LHC.}}\label{tab:fVsigma}
\end{table}

Alternatively, one could ask what luminosity is required to make a $5\sigma$ distinction between pure GF or VBF and the SM JEPs. These results are shown in Table \ref{tab:fVlumi}.

\begin{table}

\begin{tabular}{|c|cc|cc|}
\cline{2-5}
\multicolumn{1}{c}{} & \multicolumn{2}{|c}{$\mathbf{M_{jj} > 500}$ \bf{GeV}}& \multicolumn{2}{|c|}{$\mathbf{M_{jj}>250}$ \bf{GeV}}\\[0.02cm]
\hline
\textsf{5}$\sigma$ & GF & VBF & GF & VBF \\[0.05cm]
\hline
Lum [fb$^{-1}$] & 33  & 100 & 27  & 43\\
\hline
\end{tabular}
\caption{{\small Integrated luminosity required to distinguish SM from pure GF or VBF event samples at the $5\sigma$ level.}}
\label{tab:fVlumi}
\end{table}

It would seem that using a lower invariant mass cut of $250$ GeV leads to better discrimination between SM and the pure GF or pure VBF hypotheses. However, this is not necessarily the case. Lowering the invariant mass cut leads to increased statistics, which decreases the error bars on the JEPs, but it also leads to contamination from background. We will see how including the background affects our discriminating power in the next section.

\section{Effect of Background}
\label{sec:background}

 So far we have neglected the effect of contamination from background on our results. A precise estimate of this effect needs accurate simulations and/or fits to 14 TeV LHC data and is beyond the scope of this study. Nevertheless, we will estimate  the effect of the background on the discriminating power of our technique.

We can safely assume that the background JEP, $\psi_B (r)$, can be reconstructed from the data and that the Higgs signal profile, $\psi_S (r)$, can be obtained from the observed profile, $\psi_{obs} (r)$, by subtracting the background contribution:
\begin{equation}
\psi_S (r) = \psi_{obs} (r) +\frac{B}{S}\, (\psi_{obs} (r) - \psi_B (r)) \ ,
\label{eq:bkdprofile}
\end{equation}
 where $B/S$ is the background-to-signal ratio. Note that we are once again neglecting the $p_T$ dependence of the JEPs.

 This subtraction procedure introduces a correction to the statistical errors on the signal profiles derived in Sec. \ref{sec:procedure}. We make a conservative assumption that the error on background is the same size as the error on the signal for the same number of events. This introduces a scaling correction to the errors obtained in the previous section, given by a factor:
 \[
 \sqrt{1+2\frac{B}{S}} \ .
 \]

In order to get a rough estimate of the $B/S$ ratio at the 14 TeV LHC, we simulate with MadGraph the irreducible $\gamma\gamma jj$ QCD background and we apply the selection cuts of Sec. \ref{sec:KS}. Table \ref{tab:bckg} shows the results of simulations at the 8 TeV LHC, compared to the CMS results \cite{Chatrchyan:2012ufa}. Given the good agreement with the 8 TeV LHC data, we use the LO simulation to derive the signal-to-background ratio for the different selection categories at the 14 TeV LHC. Our results are shown in Table \ref{tab:bckg}. We see that the background contamination leads to an increase in the error by about 37\% using tight cuts with $M_{jj}>500$ GeV and an increase of about 52\% for the $M_{jj}>250$ GeV category.

Table \ref{tab:results-bckg} shows the corresponding ability of the 14 TeV LHC to discriminate the GF and VBF hypotheses from the SM, including the effect of background. We see that even with the inclusion of background, it is still better to consider a lower $M_{jj}$ cut of 250 GeV, to increase the discriminating power of our analysis.

\begin{table}
\begin{tabular}{|c|c|}
\multicolumn{1}{c}{ } & \multicolumn{1}{c}{\textsf{8 TeV} \sf{Background}}\\
\cline{2-2}
\multicolumn{1}{c}{} & \multicolumn{1}{|c|}{$\mathbf{M_{jj} > 500}$ \bf{GeV}} \\[0.02cm]
\hline
CMS & 0.25 fb  \\[0.05cm]
\hline
MG & 0.23 fb \\[0.05cm]
\hline
\end{tabular}
\\[0.5cm]
\begin{tabular}{|c|c|c|}
\multicolumn{1}{c}{ } & \multicolumn{2}{c}{\textsf{14 TeV} \sf{Background}}\\
\cline{2-3}
\multicolumn{1}{c}{} & \multicolumn{1}{|c}{$\mathbf{M_{jj} > 500}$ \bf{GeV}} & \multicolumn{1}{|c|}{$\mathbf{M_{jj}>250}$ \bf{GeV}}\\[0.02cm]
\hline
MG & 0.78 fb & 1.5 fb \\[0.05cm]
\hline
$S/B$ & 2.3 & 1.5\\[0.05cm]
\hline
\end{tabular}
\caption{{\small Upper Table: Background cross-sections extracted from the number of background events using 5.3 fb$^{-1}$ of data from CMS (listed in Table 2 of \cite{Chatrchyan:2012ufa}) and the MadGraph (MG) prediction for the irreducible $\gamma \gamma jj$ QCD background after tight selection cuts and after applying an additional cut, $124<m_{\gamma \gamma}<126$ GeV. Lower Table: Estimated background cross-sections and signal-background ratios at the 14 TeV LHC. }}
\label{tab:bckg}
\end{table}

\begin{table}

\begin{tabular}{|c|cc|cc|}
\cline{2-5}
\multicolumn{1}{c}{} & \multicolumn{2}{|c}{$\mathbf{M_{jj} > 500}$ \bf{GeV}}& \multicolumn{2}{|c|}{$\mathbf{M_{jj}>250}$ \bf{GeV}}\\[0.02cm]
\hline
\textsf{100 fb}$\mathsf{^{-1}}$ & GF & VBF & GF & VBF \\[0.05cm]
\hline
$\sigma$ level & 6.4  & 3.6 & 6.4  & 5.0\\
\hline
\end{tabular}
\\[0.5cm]
\begin{tabular}{|c|cc|cc|}
\cline{2-5}
\multicolumn{1}{c}{} & \multicolumn{2}{|c}{$\mathbf{M_{jj} > 500}$ \bf{GeV}}& \multicolumn{2}{|c|}{$\mathbf{M_{jj}>250}$ \bf{GeV}}\\[0.02cm]
\hline
\textsf{5}$\sigma$ & GF & VBF & GF & VBF \\[0.05cm]
\hline
Lum [fb$^{-1}$] & 61  &  190 &  61  & 100\\
\hline
\end{tabular}
\caption{{\small Upper Table: Expected $\sigma-$level distinction between SM and pure GF/VBF event samples using 100 fb$^{-1}$ of luminosity at the 14 TeV LHC including the estimated effect of background. Lower Table: Integrated luminosity required to distinguish SM from pure GF/VBF event samples at the $5\sigma$ level after subtracting the background JEP.}}
\label{tab:results-bckg}
\end{table}

\section{Summary and Conclusions}
\label{sec:conclusions}
Separation of Higgs production modes is important in order to directly measure the Higgs couplings to gluons and vector bosons. Conventional methods of separating the VBF contribution to Higgs production use cuts on Higgs boson events with two forward jets. However, a sizeable fraction of GF events still remains in this sample. Kinematic discriminators provide only a modest ability to further separate GF from VBF events.

In this paper, we made the observation that the jets associated with VBF are initiated by quarks whereas the jets associated with GF are dominantly initiated by gluons. We then presented a new tool for discriminating Higgs boson production mechanisms based on the analysis of the JEP associated with the central jet in $H + 2 \textrm{ jets}$ events. We used the profiles to construct a discriminating variable, $f_V$, which can be regarded as the fraction of VBF events in a given sample. We constructed two test scenarios, where the Higgs is produced via pure VBF and pure GF, and used these to benchmark our ability to discriminate against the SM.

We found the central value for $f_V$ using a theoretical pQCD calculation of the JEP (Table \ref{tab:fVerrors}). We estimated the expected errors on the measurement of $f_V$ at the 14 TeV run of the LHC for different luminosities using Pythia simulations, including the effect of background.

The main results of our paper are shown in Table \ref{tab:results-bckg}, where we show how well the pure GF and pure VBF hypotheses can be separated from the SM by using the measured values of $f_V$. We find that with a $100$ fb$^{-1}$ of luminosity at the 14 TeV LHC, both the pure GF and via pure VBF hypotheses can be excluded at the $5\sigma$ level. Our method should be included in a global analysis to further strengthen its discriminatory power.

The use of JEPs to separate various operators contributing to Higgs production is the novel feature of this work. A similar technique can be applied to probe new physics models. A couple of applications are:  
  \begin{itemize}
\item Separation of $QQ\chi\chi$ versus $GG\chi\chi$ contact operator coefficients in dark matter mono-jet searches.
\item Distinction between different types of dijet resonances (colorons, Z-primes, etc.).
\end{itemize}
\section{Acknowledgments}
We would like to thank J. Huston and C. Vermilion for useful discussions about SpartyJet. This work was supported by the U.S. National Science Foundation under Grant Nos. PHY-0855561 and PHY-0854889; the National Science Council of R.O.C. under Grant No. NSC-101-2112-M-001-006-MY3; and the National Natural Science Foundation of China under Grant No. NSFC11245007. V.R. would like to thank the KITP where part of this work was completed with the support of Grant No. NSF PHY11-25915.
\bibliography{bibhjj-final}

\begin{thebibliography}{18}
\expandafter\ifx\csname natexlab\endcsname\relax\def\natexlab#1{#1}\fi
\expandafter\ifx\csname bibnamefont\endcsname\relax
  \def\bibnamefont#1{#1}\fi
\expandafter\ifx\csname bibfnamefont\endcsname\relax
  \def\bibfnamefont#1{#1}\fi
\expandafter\ifx\csname citenamefont\endcsname\relax
  \def\citenamefont#1{#1}\fi
\expandafter\ifx\csname url\endcsname\relax
  \def\url#1{\texttt{#1}}\fi
\expandafter\ifx\csname urlprefix\endcsname\relax\def\urlprefix{URL }\fi
\providecommand{\bibinfo}[2]{#2}
\providecommand{\eprint}[2][]{\url{#2}}

\bibitem[{\citenamefont{Aad et~al.}(2012)}]{Aad:2012tfa}
\bibinfo{author}{\bibfnamefont{G.}~\bibnamefont{Aad}} \bibnamefont{et~al.}
  (\bibinfo{collaboration}{ATLAS Collaboration}), \bibinfo{journal}{Phys.Lett.}
  \textbf{\bibinfo{volume}{B716}}, \bibinfo{pages}{1} (\bibinfo{year}{2012}),
  \eprint{hep-ex/1207.7214}.

\bibitem[{\citenamefont{Chatrchyan et~al.}(2012)}]{Chatrchyan:2012ufa}
\bibinfo{author}{\bibfnamefont{S.}~\bibnamefont{Chatrchyan}}
  \bibnamefont{et~al.} (\bibinfo{collaboration}{CMS Collaboration}),
  \bibinfo{journal}{Phys.Lett.} \textbf{\bibinfo{volume}{B716}},
  \bibinfo{pages}{30} (\bibinfo{year}{2012}), \eprint{hep-ex/1207.7235}.

\bibitem[{ATL(2013)}]{ATLAS-CONF-2013-034}
\bibinfo{type}{Tech. Rep.} \bibinfo{number}{ATLAS-CONF-2013-034},
  \bibinfo{institution}{CERN}, \bibinfo{address}{Geneva}
  (\bibinfo{year}{2013}).

\bibitem[{CMS(2013)}]{CMS-PAS-HIG-13-005}
\bibinfo{type}{Tech. Rep.} \bibinfo{number}{CMS-PAS-HIG-13-005},
  \bibinfo{institution}{CERN}, \bibinfo{address}{Geneva}
  (\bibinfo{year}{2013}).

\bibitem[{H-g()}]{H-global-fit}
\bibinfo{note}{See, for example, J. Ellis and T. You (2013), hep-ph/1303.3879,
  and references therein.}

\bibitem[{\citenamefont{Acosta et~al.}(2005)}]{Acosta:2005ix}
\bibinfo{author}{\bibfnamefont{D.}~\bibnamefont{Acosta}} \bibnamefont{et~al.}
  (\bibinfo{collaboration}{CDF Collaboration}), \bibinfo{journal}{Phys.Rev.}
  \textbf{\bibinfo{volume}{D71}}, \bibinfo{pages}{112002}
  (\bibinfo{year}{2005}), \eprint{hep-ex/0505013}.

\bibitem[{\citenamefont{Aad et~al.}(2011)}]{Aad:2011kq}
\bibinfo{author}{\bibfnamefont{G.}~\bibnamefont{Aad}} \bibnamefont{et~al.}
  (\bibinfo{collaboration}{Atlas Collaboration}), \bibinfo{journal}{Phys.Rev.}
  \textbf{\bibinfo{volume}{D83}}, \bibinfo{pages}{052003}
  (\bibinfo{year}{2011}), \eprint{hep-ex/1101.0070}.

\bibitem[{CMS(2010)}]{CMS-PAS-QCD-10-013}
\bibinfo{type}{Tech. Rep.} \bibinfo{number}{CMS-PAS-QCD-10-013},
  \bibinfo{institution}{CERN}, \bibinfo{address}{2010. Geneva}
  (\bibinfo{year}{2010}).

\bibitem[{\citenamefont{Sjostrand et~al.}(2008)\citenamefont{Sjostrand, Mrenna,
  and Skands}}]{Sjostrand:2007gs}
\bibinfo{author}{\bibfnamefont{T.}~\bibnamefont{Sjostrand}},
  \bibinfo{author}{\bibfnamefont{S.}~\bibnamefont{Mrenna}}, \bibnamefont{and}
  \bibinfo{author}{\bibfnamefont{P.~Z.} \bibnamefont{Skands}},
  \bibinfo{journal}{Comput.Phys.Commun.} \textbf{\bibinfo{volume}{178}},
  \bibinfo{pages}{852} (\bibinfo{year}{2008}), \eprint{hep-ph/0710.3820}.

\bibitem[{pyt()}]{pythiaA}
\bibinfo{note}{PYTHIA-Tune A. Monte Carlo samples are generated using the
  following tuned parameters in PYTHIA: PARP(67) = 4.0, MSTP(82) = 4, PARP(82)
  = 2.0, PARP(84) = 0.4, PARP(85) = 0.9, PARP(86) = 0.95, PARP(89) = 1800.0,
  PARP(90) = 0.25}.

\bibitem[{\citenamefont{Li et~al.}(2011)\citenamefont{Li, Li, and
  Yuan}}]{Li:2011hy}
\bibinfo{author}{\bibfnamefont{H.-n.} \bibnamefont{Li}},
  \bibinfo{author}{\bibfnamefont{Z.}~\bibnamefont{Li}}, \bibnamefont{and}
  \bibinfo{author}{\bibfnamefont{C.-P.} \bibnamefont{Yuan}},
  \bibinfo{journal}{Phys.Rev.Lett.} \textbf{\bibinfo{volume}{107}},
  \bibinfo{pages}{152001} (\bibinfo{year}{2011}), \eprint{hep-ph/1107.4535}.

\bibitem[{\citenamefont{Li et~al.}(2012)\citenamefont{Li, Li, and
  Yuan}}]{Li:2012bw}
\bibinfo{author}{\bibfnamefont{H.-n.} \bibnamefont{Li}},
  \bibinfo{author}{\bibfnamefont{Z.}~\bibnamefont{Li}}, \bibnamefont{and}
  \bibinfo{author}{\bibfnamefont{C.-P.} \bibnamefont{Yuan}},
  \bibinfo{journal}{Phys.Rev.} \textbf{\bibinfo{volume}{D87}},
  \bibinfo{pages}{074025} (\bibinfo{year}{2012}), \eprint{hep-ph/1206.1344}.

\bibitem[{\citenamefont{Cacciari et~al.}(2008)\citenamefont{Cacciari, Salam,
  and Soyez}}]{Cacciari:2008gp}
\bibinfo{author}{\bibfnamefont{M.}~\bibnamefont{Cacciari}},
  \bibinfo{author}{\bibfnamefont{G.~P.} \bibnamefont{Salam}}, \bibnamefont{and}
  \bibinfo{author}{\bibfnamefont{G.}~\bibnamefont{Soyez}},
  \bibinfo{journal}{JHEP} \textbf{\bibinfo{volume}{0804}}, \bibinfo{pages}{063}
  (\bibinfo{year}{2008}), \eprint{hep-ph/0802.1189}.

\bibitem[{\citenamefont{Alwall et~al.}(2011)\citenamefont{Alwall, Herquet,
  Maltoni, Mattelaer, and Stelzer}}]{Alwall:2011uj}
\bibinfo{author}{\bibfnamefont{J.}~\bibnamefont{Alwall}},
  \bibinfo{author}{\bibfnamefont{M.}~\bibnamefont{Herquet}},
  \bibinfo{author}{\bibfnamefont{F.}~\bibnamefont{Maltoni}},
  \bibinfo{author}{\bibfnamefont{O.}~\bibnamefont{Mattelaer}},
  \bibnamefont{and} \bibinfo{author}{\bibfnamefont{T.}~\bibnamefont{Stelzer}},
  \bibinfo{journal}{JHEP} \textbf{\bibinfo{volume}{1106}}, \bibinfo{pages}{128}
  (\bibinfo{year}{2011}), \eprint{hep-ph/1106.0522}.

\bibitem[{\citenamefont{Mangano et~al.}(2007)\citenamefont{Mangano, Moretti,
  Piccinini, and Treccani}}]{Mangano:2006rw}
\bibinfo{author}{\bibfnamefont{M.~L.} \bibnamefont{Mangano}},
  \bibinfo{author}{\bibfnamefont{M.}~\bibnamefont{Moretti}},
  \bibinfo{author}{\bibfnamefont{F.}~\bibnamefont{Piccinini}},
  \bibnamefont{and} \bibinfo{author}{\bibfnamefont{M.}~\bibnamefont{Treccani}},
  \bibinfo{journal}{JHEP} \textbf{\bibinfo{volume}{0701}}, \bibinfo{pages}{013}
  (\bibinfo{year}{2007}), \eprint{hep-ph/0611129}.

\bibitem[{\citenamefont{Alwall et~al.}(2008)\citenamefont{Alwall, Hoche,
  Krauss, Lavesson, Lonnblad et~al.}}]{Alwall:2007fs}
\bibinfo{author}{\bibfnamefont{J.}~\bibnamefont{Alwall}},
  \bibinfo{author}{\bibfnamefont{S.}~\bibnamefont{Hoche}},
  \bibinfo{author}{\bibfnamefont{F.}~\bibnamefont{Krauss}},
  \bibinfo{author}{\bibfnamefont{N.}~\bibnamefont{Lavesson}},
  \bibinfo{author}{\bibfnamefont{L.}~\bibnamefont{Lonnblad}},
  \bibnamefont{et~al.}, \bibinfo{journal}{Eur.Phys.J.}
  \textbf{\bibinfo{volume}{C53}}, \bibinfo{pages}{473} (\bibinfo{year}{2008}),
  \eprint{hep-ph/0706.2569}.

\bibitem[{\citenamefont{Delsart et~al.}(2012)\citenamefont{Delsart, Geerlings,
  Huston, Martin, and Vermilion}}]{Delsart:2012jm}
\bibinfo{author}{\bibfnamefont{P.-A.} \bibnamefont{Delsart}},
  \bibinfo{author}{\bibfnamefont{K.~L.} \bibnamefont{Geerlings}},
  \bibinfo{author}{\bibfnamefont{J.}~\bibnamefont{Huston}},
  \bibinfo{author}{\bibfnamefont{B.~T.} \bibnamefont{Martin}},
  \bibnamefont{and} \bibinfo{author}{\bibfnamefont{C.~K.}
  \bibnamefont{Vermilion}} (\bibinfo{year}{2012}), \eprint{hep-ex/1201.3617}.

\bibitem[{\citenamefont{Cacciari et~al.}(2012)\citenamefont{Cacciari, Salam,
  and Soyez}}]{Cacciari:2011ma}
\bibinfo{author}{\bibfnamefont{M.}~\bibnamefont{Cacciari}},
  \bibinfo{author}{\bibfnamefont{G.~P.} \bibnamefont{Salam}}, \bibnamefont{and}
  \bibinfo{author}{\bibfnamefont{G.}~\bibnamefont{Soyez}},
  \bibinfo{journal}{Eur.Phys.J.} \textbf{\bibinfo{volume}{C72}},
  \bibinfo{pages}{1896} (\bibinfo{year}{2012}), \eprint{hep-ph/1111.6097}.

\end{thebibliography}

\end{document}